\documentclass[twocolumn,showpacs,preprintnumbers,amssymb]{revtex4}
\usepackage{graphicx}
\usepackage{dcolumn}
\usepackage{bm}

\begin{document}
\preprint{CCOC-02-04}
\title{Generalized Phase Synchronization in unidirectionally coupled chaotic oscillators} 
\author{Dae-Sic Lee} 
\author{Won-Ho Kye} 
\email{whkye@phys.paichai.ac.kr}
\author{Sunghwan Rim}
\author{Tae-Yoon Kwon}
\author{Chil-Min Kim}
\email{chmkim@mail.paichai.ac.kr}
\affiliation{National Creative Research Initiative Center for Controlling Optical Chaos,
Pai-Chai University, Daejeon 302-735, Korea}
\begin{abstract}
	  We investigate phase synchronization between two identical or detuned response
	  oscillators coupled to a slightly different drive oscillator. Our
          result is that phase synchronization can occur between response oscillators 
	  when they are driven by correlated (but not
          identical) inputs from the drive oscillator. We call this phenomenon
          Generalized Phase Synchronization (GPS) and clarify its characteristics
          using Lyapunov exponents and phase difference plots.
\end{abstract}

\maketitle

Synchronization has been of much interest 
since Huygens' first description of it in two pendulum clocks on a wall \cite{Huygens}.
The report that synchronization can be observed even in chaotic systems 
gave new rise to scientific attention to the phenomenon \cite{SyncOrg0,SyncOrg1}.
Over the past decade, synchronization in coupled chaotic oscillators 
has been intensely investigated for the understanding 
of its fundamental role in coupled nonlinear systems 
and the possibility of applications in various fields \cite{PSExp,Circuit,NeuronSync,Secure,LaserPS}. 
What characterizes synchronization is the convergence of the distance between the state variables
of drive and response systems to zero due to weak interaction.
Several different types of synchronization 
i.e., phase synchronization (PS) \cite{PhaseSync,Lee,Type-I}, lag synchronization (LS) \cite{Lag}, 
complete synchronization (CS) \cite{SyncOrg1}, and generalized synchronization (GS) \cite{GenSync}
have been observed in coupled chaotic systems.

While CS, PS, and LS are observed in identical or slightly detuned 
systems (due to parameter mismatch) \cite{SyncOrg0,SyncOrg1,PhaseSync}, GS is
observed in coupled oscillators with different dynamics \cite{GenSync}.
When chaotic signals of a drive oscillator 
are fed into response oscillators, 
above a critical coupling
the response oscillators lose their exponential 
instability in the transverse direction and their state variables 
converge to the same value.
This convergence is the main character of GS.
Since the attractors of response oscillators converge to the same image 
in the GS regime, GS implies the emergence of a functional 
relation between drive and response oscillators such that 
${\bf x}_1={\bf H}({\bf x}_2)$ \cite{GenSync}, where $ {\bf x}_1 $ and ${\bf x}_2$ are
the state vectors of drive and response oscillators, respectively. 

In mutually coupled chaotic oscillators,
there have been extensive investigations on the whole synchronization phenomena
(PS, LS and CS) and various transition scenarios clarified \cite{PhaseSync,Lag}.
Most of the investigations in unidirectionally coupled chaotic oscillators 
have been concentrated on GS transition and its applications.
Also, some trials have been made to unveil the relation between PS and GS: 
Parlitz et al. \cite{Palitz} experimentally studied PS in unidirectionally coupled analog computer
and Zheng et al. \cite{Hu} theoretically demonstrated that GS can be
weaker than PS depending on parameter mismatchs.
Nevertheless, there remain unclarified questions concerning to 
PS in unidirectionally coupled chaotic oscillators.
Particular questions are {\it how PS phenomena established in the drive-response and  
the response-response  systems differ from each other} and {\it how they develop to GS}. 

In this report, we investigate transition to PS in unidirectionally coupled
chaotic oscillators driven by two different types of chaotic signal.
We demonstrate that PS in the response-response system is  
induced by PS in the drive-response system when the driving
signals are identical. 
However, we find that when correlated but not identical driving signals are used, 
PS is established in the response-response system but not in the drive-response system. 
We call this special PS phenomenon {\it generalized phase synchronization} (GPS)
and discuss its relation with GS.

How to define the phase for a chaotic system is an important issue and 
an active field of investigation in nonlinear dynamics.
So far several methods, e.g. using phase space projection \cite{Spline,PhaseSync}, 
Hilbert transformation\cite{LaserPS}, and wavelet transformation \cite{Wavelet} etc.,  
have been suggested and there have been done extensive investigation based on these methods.
The method using phase space projection is the most simple one 
to define the phase in R\"ossler oscillator
and it enables us to use analytic treatment in analyzing the phase dynamics.
We follow this approach.  
So the phase is defined by the simple geometric function: 
$\theta_i=\arctan(y_i/x_i)$, where $i=1$ for a drive oscillator 
and $i=2,3$ for response oscillators. 

To demonstrate the conventional PS phenomenon established in the drive-response, 
we consider the unidirectionally coupled R\"ossler 
oscillators with slight parameter mismatch (see the caption of Fig. 1).
The identical signal $y_1$ from drive oscillator is fed into two responses and
the phase difference between drive and response oscillators
can be written as follows \cite{PhaseSync,Lee}:
\begin{eqnarray}
\dot{\phi}_{1k}&=&\Delta \omega -\frac{\epsilon}{2}\frac{R_1}{R_k} \sin\phi_{1k}+\xi(t),
\end{eqnarray}
where $\phi_{1k}= \theta_1-\theta_k$ and  $R_k=\sqrt{x_k^2+y_k^2}$ and $k=2,3$.
Here, $\Delta \omega$ is the frequency mismatch between drive and response
oscillators and $\xi(t)$ is the fast fluctuating term  
which plays the role of effective noise.
It is known that the above dynamics is governed by 
Type-I intermittency in the presence of noise and
that PS is established when the channel width is deeper 
than the maximum of the effective noise $\xi(t)$ \cite{Type-I}. 
Accordingly, we can estimate the onset point of PS by considering the fixed point condition: $\dot{\phi}_{1k}= 0$.
It leads to the equation: $\phi_{1k}=\arcsin(2\Delta\omega R_k/\epsilon R_1) $ 
where we ignore the effective noise term $\xi$
because it is negligible in the PS regime \cite{PhaseSync}.  
Accordingly the onset solution of the fixed point is $\phi^*_{1k}=\pi/2$ when  $ 2 \Delta \omega R_k/\epsilon R_1= 1$.
Thus the critical coupling can be estimated by  $\epsilon\approx 2 \Delta \omega$ since $R_1/R_k\approx 1$ 
between slightly detuned systems \cite{PhaseSync}.     
Then the critical coupling for PS can be estimated by 
$\epsilon_c=2 \Delta \omega= 2 (\omega_d -\omega_r)=0.03$.

Figure 1 shows the stroboscopic phase trajectories and 
probability distributions $P(\theta_{1,2})$ 
of chaotic oscillators 1 and 2 just above the critical point 
with the reference oscillator (see Ref. \cite{Strobo} for our definition). 
One can see that PS occurs between oscillators 1 and 3 (Fig. 1 (a) and (b)) 
as well as between oscillators 2 and 3 (Fig. 1 (c) and (d)) as 
oscillators 1 and 2 take the preference directions 
(i.e., $\langle \exp(i \theta) \rangle=\int \exp(i \theta) P(\theta) d \theta  \neq 0 $). 
This implies that the rotational symmetry  
on the projective attractor is broken due to PS transition, which
is the indisputable evidence of PS \cite{PhaseSync,Type-I,Lee}. 
Accordingly, we understand that PS established in the response-response system
is due to PS in the drive-response system, i.e., above the critical coupling  
$\dot{\phi}_{12}=0$ and $\dot{\phi}_{13}=0$ imply $\dot{\phi}_{23}=0$ in Eq. (1). 
In other words, PS in the drive-response system coincides 
with PS in the response-response system when identical driving signals are used. 
 
\begin{figure}
\begin{center}
\rotatebox[origin=c]{0}{\includegraphics[width=8.3cm]{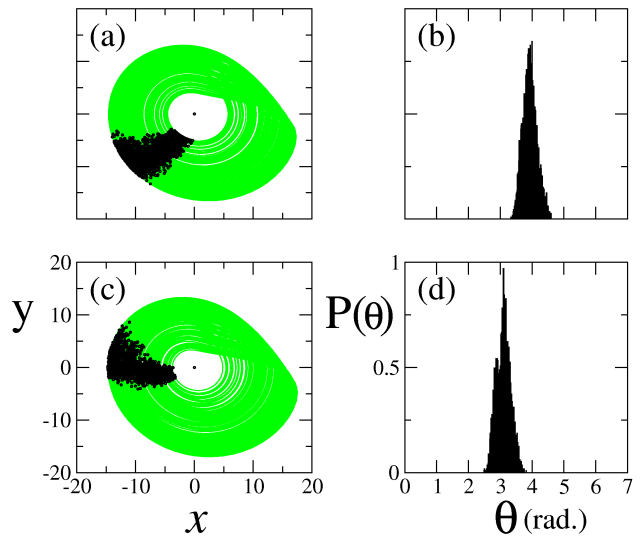}}
\caption{PS at $\epsilon=0.04$ in coupled R\"ossler oscillators: 
                $\dot{x}_1=-\omega_d y_1-z_1$, 
                $\dot{y}_1=\omega_d x_1 + 0.15 y_1$,
                $\dot{z}_1=0.2 + z_1(x_1-10)$, 
                $\dot{x}_{2,3}=-\omega_r y_{2,3}-z_{2,3}$,
                $\dot{y}_{2,3}=\omega_r x_{2,3} + 0.165 y_{2,3} +\epsilon(y_1-y_{2,3})$,
                $\dot{z}_{2,3}=0.2 + z_{2,3}(x_{2,3}-10)$,
	where $\omega_d=1.015$ and $\omega_r=1.0$.
	(a) stroboscopic phase trajectory (black dots) of oscillator 1  with reference oscillator 3 \cite{Strobo}.
	Gray dots show the whole attractor of oscillator 1 without stroboscopic sampling. 
	(c) stroboscopic phase trajectory (black dots) of oscillator 2  with reference oscillator 3.  
	(b) and (d) probability distributions of (a) and (c), respectively.
}
\end{center}
\end{figure}

\begin{figure}
\begin{center}
\rotatebox[origin=c]{0}{\includegraphics[width=8.3cm]{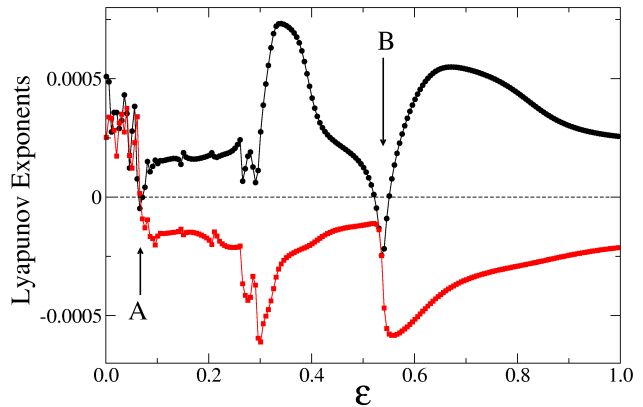}}
\caption{Two largest transverse Lyapunov exponents when $\omega_d=0.7$ and $\omega_{r}=1.0$. 
There are two transition points, A and B.}
\end{center}
\end{figure}
Next, we consider two response R\"ossler oscillators driven 
by the correlated signals, $x_1$ and $y_1$,  instead of identical ones: 
\begin{eqnarray}
		\dot{x}_1&=&-\omega_d y_1-z_1, \nonumber\\
		\dot{y}_1&=&\omega_d x_1 + 0.165 y_1, \nonumber\\
		\dot{z}_1&=&0.2 + z_1(x_1-10),      \\
		\nonumber\\
		\dot{x}_2&=&-\omega_{r} y_2-z_2 +\epsilon(x_1-x_2), \nonumber\\
		\dot{y}_2&=&\omega_{r} x_2 + 0.165 y_2, \nonumber\\
		\dot{z}_2&=&0.2 + z_2(x_2-10),      \\
		\nonumber\\
		\dot{x}_3&=&-\omega_{r} y_3-z_3, \nonumber \\
		\dot{y}_3&=&\omega_{r} x_3 + 0.165 y_3 +\epsilon(y_1-y_3), \nonumber \\
		\dot{z}_3&=&0.2 + z_3(x_3-10), 
\end{eqnarray} 
where the correlated signal $x_1$ and $y_1$ of oscillator 1 are 
fed into oscillators 2 and 3, respectively.
In real systems, noise and delay in propagating channel are unavoidable.
Thus the above system models a real situation in which
two response systems are driven by correlated signals, $x$ and $x^\prime$ 
where $x^\prime$ is a distorted version of $x$.
We propose the above system for studying PS in unidirectionally coupled systems
and its relation to GS.

The difference dynamics between two response oscillators is 
given by $\Delta \dot{\bf X}= A \Delta \bf{X} + {\Xi}(t)$
where $\Delta {\bf X}= {\bf x_2}-{\bf x_3}$, $A=((0, -\omega_r, -1),~(\omega_r, 0.165, 0),~(0, 0, -10))$, and 
${\bf \Xi}= \mbox{diag}(\epsilon(x_1-x_2), ~ -\epsilon(y_1-y_3), ~ z_2 x_2 - z_3 x_3)$.
By iterating this dynamics, we can find transverse Lyapunov exponents
describing the relative motion of oscillators 2 and 3.
In Fig. 2, we see two transition points, A and B, which, as we will see below,
correspond to two different types of PS:
namely, GPS at A and the conventional PS at B.

\begin{figure}
\begin{center}
\rotatebox[origin=c]{0}{\includegraphics[width=8.3cm]{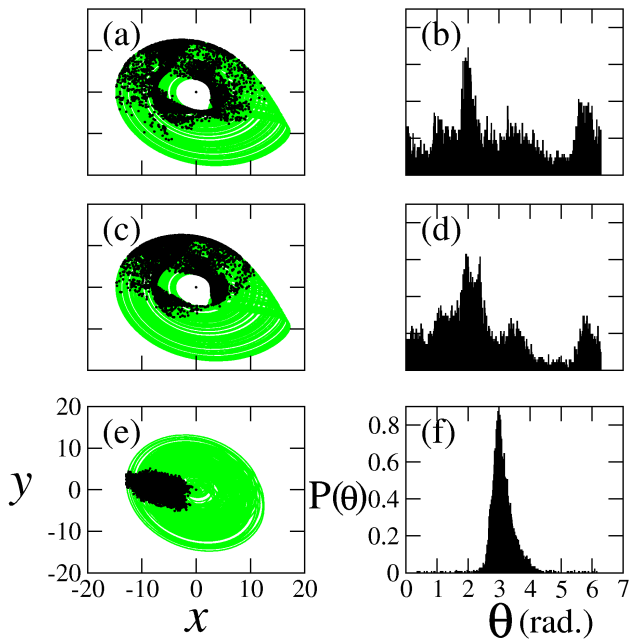}}
\caption{
	GPS in Eq. (2)-(4)  at $\epsilon=0.2$:
	(a) stroboscopic phase trajectory (black dots) of oscillator 1 with oscillator 3 \cite{Strobo};
	(c) oscillator 1 with oscillator 2;
	(e) oscillator 3 with oscillator 2; (b) (d) (f) probability distributions of (a), (c), and (e), respectively.}
\end{center}
\end{figure}

The phase difference between drive and response oscillators is given by   
\begin{eqnarray}
	\dot{\phi}_{1k}&=& \Delta \omega -B(\theta_1, \theta_k) \sin\phi_{1k} 
		+\eta_k(\theta_1, \theta_k),
\end{eqnarray}
where, $\Delta\omega=\omega_d-\omega_{r}$,
\begin{eqnarray}
B(\theta_1, \theta_k)&=&\frac{\epsilon}{2}\frac{R_1}{R_k}-0.15 \cos(\theta_1+\theta_k), \nonumber\\
\eta_k(\theta_1, \theta_k)&=& \frac{\epsilon(2 k -5)}{2}\frac{R_1}{R_k}\sin(\theta_1+\theta_k) \nonumber\\
			& &-(0.015-\epsilon)\sin\theta_k\cos\theta_k \nonumber\\
			& &+(\frac{z_1}{R_1}\sin\theta_1 -\frac{z_k}{R_k}\sin\theta_k). \nonumber
\end{eqnarray}
We can see a $k$-dependent term in $\eta_k(\theta_1, \theta_k)$ 
which is due to the driving by correlated signals from the drive oscillator in Eq. (2)-(4). 
We can obtain the critical point for PS transition which is $\epsilon_c=-2\Delta \omega=0.6$, 
in accordance with the argument of the former case (below Eq. (1)). 
The drive and response oscillators develop to a PS state at this critical value 
and PS between oscillators 2 and 3 is induced above this critical coupling.
The critical value of the critical coupling agrees with that of the transition point  
indicated by point B ($\epsilon \approx 0.55$). 
Thus we understand that B corresponds to
conventional PS transition point at which the three oscillators develop to PS simultaneously.

\begin{figure}
\begin{center}
\rotatebox[origin=c]{0}{\includegraphics[width=8.3cm]{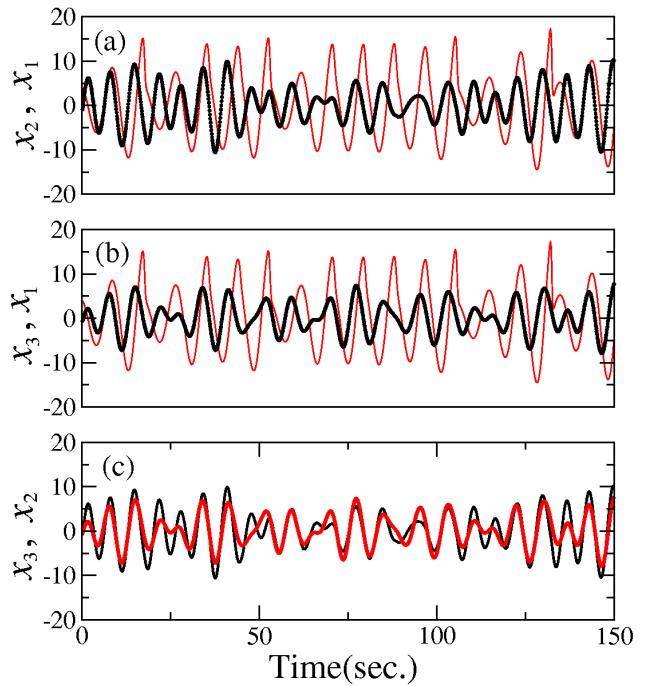}}
\caption{Temporal behaviors of drive and response oscillators in the GPS regime 
	when $\epsilon=0.2$: (a) $x_1$ and $x_2$ (b) $x_1$ and $x_3$ (c) $x_2$ and $x_3$.}
\end{center}
\end{figure}
We need to inspect the phenomenon at reference point A, 
which is described as crossing to the negative value in one of the Lyapunov exponents.    
Fig. 3 shows the trajectories in phase space and probability distributions near reference point A.
A PS state appears only in the response-response system (Fig. 3 (e) and (f)) without 
PS in the drive-response system (because probability distributions are 
not localized in Fig. 3 (a)-(d)), which is different from the case of  
the reference point B. 
We call this phenomenon GPS on the analogy of GS in which two state variables, 
${\bf x}_2$ and ${\bf x}_3$ coverge to the same value.
In GPS the phases $\theta_2$ and $\theta_3$ are bounded by a constant.  
Figure 4 shows the temporal behaviors of each oscillator at the same coupling 
strength as that of Fig. 3. 
We note that PS is established in oscillator 2 and 3 as the
phases are mostly matched in Fig. 4 (c), 
while phase slippings appear in oscillators 1 and 2 
(Fig. 4 (a)) or oscillators 1 and 3 (Fig. 4 (b)), intermittently.  
Finally, we remark that GPS is a novel phenomenon characterized by PS in the response-response
system, and thus it is different from conventional PS\cite{PhaseSync}.

In Fig. 5, we also observe that GPS appears when the responses are slightly detuned with
frequencies of $1.0\pm0.005$, and it destabilizes for frequencies of $1.0\pm0.01$.
This implies that GPS is a real phenomenon that should be experimentally observable \cite{Remark}.
And the attractor deformations observed in Fig. 3 (e) and Fig. 5 (e) seem to 
be originated from the driving signal of different natural frequency.
The similar phenomenon is often observed in unidirectionally
coupled systems with different dynamics \cite{GenSync}. 
It was shown recently \cite{PhaseSync,Lee} that the phase defined by geometrical function like ours 
and the phase based on the Hilbert transformation practically coincide. 
So we think our result is independent of the method of defining the phase.

\begin{figure}
\begin{center}
\rotatebox[origin=c]{0}{\includegraphics[width=8.3cm]{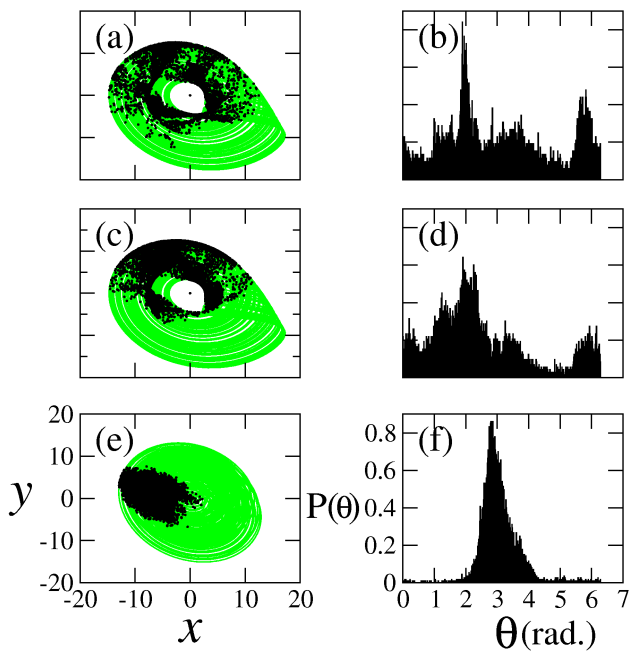}}
\caption{Appearance of GPS when two response oscillators are slightly
	 detuned. The natural frequencies of response
	 oscillators are $w_{r}=1.005$ (of oscillator 2) and  $w_{r}=0.995$ (of oscillator 3), respectively.
	 Others parameters are those of Fig. 3.}
\end{center}
\end{figure}

In conclusion, we have studied PS in unidirectionally 
coupled chaotic systems with parameter mismatch.
And we have focusedly clarified the relationship of PS phenomena 
in the drive-response and PS in the response-response systems.
When the driving signals are identical, PS in the drive-response system 
corresponds to PS in the response-response system and the system develops to GS as 
the coupling strength increases: PS$\rightarrow $GS.
When the driving signals are correlated (but not identical), 
PS is established in the response-response system
but not the drive-response system. We call this phenomenon GPS.
The GPS state transits to PS when coupling strenght increases.
The results are confirmed by the analysis of Lyapunov exponents, 
phase trajectories, and time series.     
We expect that the GPS concept could be used for analyzing weak interdependences 
of data coming from weakly correlated systems such as neuronal systems \cite{NeuronSync}, 
cardiac oscillators \cite{Card}, and ecological systems \cite{eco} etc. 

The authors thank J. Kurths, S.-Y. Lee, and M.S. Kurdoglyan for helpful discussions.
This work is supported by Creative Research Initiatives of the Korean Ministry of Science and Technology.


\begin{thebibliography}{150}
\bibitem{Strobo} 
		One can collect the values of projective coordinate $(x_t, y_t)$ or phase value $\theta_t$
		of a target oscillator, whenever a reference oscillator
		crosses a Poincar\'e section, e.g., we take $x_r < 0, y_r=0$.
		In this case we say that the plot of the values of $(x_t,y_t)$ is the stroboscopic
		trajectory in phase space with the reference oscillator.
		$\theta_t$ is the corresponding phase. 
		The probability distribution $P(\theta_t)$ is the normalized histogram
		of the stroboscopic phases.
\bibitem{Remark} On the other hand, we could not observe GPS regime in the case  $\omega_d > \omega_r$.
\end{thebibliography}
\end{document}